\documentclass[twocolumn,preprintnumbers,superscriptaddress,amssymb,apl,aip,10pt]{revtex4-1}

\usepackage{graphicx}
\usepackage{dcolumn}
\usepackage{bm}
\usepackage{color}
\usepackage{amsmath,amsfonts,amssymb}  
\usepackage{graphicx}
\usepackage[english]{babel}
\usepackage{booktabs,longtable}
\usepackage{natbib}

\def\gsim{\mathrel{\rlap{\lower3pt\hbox{$\sim$}}\raise1pt\hbox{$>$}}}

\begin{document}

\author{J.~Sampaio}
\altaffiliation[Present address: ]{Laboratoire de Physique des Solides, Univ. Paris-Sud, Universit\'e Paris-Saclay, CNRS UMR 8502, 91405 Orsay Cedex, France}
\affiliation{Unit\'e Mixte de Physique, CNRS, Thales, Univ. Paris-Sud, Universit\'e Paris-Saclay, 91767, Palaiseau, France}
\author{A.~V.~Khvalkovskiy}
\affiliation{Samsung Electronics, Semiconductor R\&D Center (Grandis), San Jose, CA 95134, USA}
\affiliation{ Moscow Institute of Physics and Technology, State University, Moscow, Russia }
\author{M.~Kuteifan}
\affiliation{Department of Electrical and Computer Engineering, University of California at San Diego, La Jolla, CA 92093-0407, USA}
\author{M.~Cubukcu}
\affiliation{Unit\'e Mixte de Physique, CNRS, Thales, Univ. Paris-Sud, Universit\'e Paris-Saclay, 91767, Palaiseau, France}
\author{D.~Apalkov}
\affiliation{Samsung Electronics, Semiconductor R\&D Center (Grandis), San Jose, CA 95134, USA}
\author{V.~Lomakin}
\affiliation{Department of Electrical and Computer Engineering, University of California at San Diego, La Jolla, CA 92093-0407, USA}
\author{V.~Cros}
\affiliation{Unit\'e Mixte de Physique, CNRS, Thales, Univ. Paris-Sud, Universit\'e Paris-Saclay, 91767, Palaiseau, France}
\author{N.~Reyren}
\email{nicolas.reyren@thalesgroup.com}
\affiliation{Unit\'e Mixte de Physique, CNRS, Thales, Univ. Paris-Sud, Universit\'e Paris-Saclay, 91767, Palaiseau, France}

\title{Disruptive effect of Dzyaloshinskii-Moriya interaction on  the MRAM cell performance}

\date{\today}

\begin{abstract}
In order to increase the thermal stability of a magnetic random access memory (MRAM) cell, materials with high spin-orbit interaction are often introduced in the storage layer. As a side effect, a strong Dzyaloshinskii-Moriya interaction (DMI) may arise in such systems. Here we investigate the impact of DMI on the magnetic cell performance, using micromagnetic simulations. We find that DMI strongly promotes non-uniform magnetization states and non-uniform switching modes of the magnetic layer. It appears to be detrimental for both the thermal stability of the cell and its switching current, leading to considerable deterioration of the cell performance even for a moderate DMI amplitude. 
\end{abstract}

\maketitle

Recently, development of magnetic random access memories (MRAM) for dense memory products such as DRAM and SRAM became focused on magnetic cells with a high perpendicular magnetic anisotropy (PMA). These designs are believed to offer an improved thermal stability at very advanced technological nodes of $20$\,nm and below \cite{Gajek2012,Khvalkovskiy2013}. The PMA storage (a.k.a. `free') magnetic layer is based on magnetically soft CoFeB, which has a good lattice matching with the MgO barrier. The interface between MgO and CoFeB provides sufficiently strong PMA to hold perpendicular a CoFeB layer of about 1\,nm.\cite{Ikeda2010} In order to further enhance the thermal stability of the cell, elements with a strong spin-orbit coupling (SOC), such as W, Pt, Ta or Ir are often introduced into the free layer (FL) \cite{Yuasa2010,HOhno2012,HOhno2013,HOhno2014}. However recent studies demonstrated that a very large Dzyaloshinskii-Moriya exchange interaction (up to a large fraction of the Heisenberg exchange) may arise at the FM/SOC film interface \cite{Heinze2011, MoreauLuchaire}. DMI can dramatically change the magnetic state of the film. It was shown to induce a significant spin tilt at the borders \cite{Sampaio2013,Rohart2013}, for large DMI amplitude it can stabilize cycloidal states and skyrmion lattices \cite{Nagaosa2013}. Moreover it drastically changes the domain wall (DW) energy and, thus, the magnetic switching process \cite{Pizzini2014}, both under field and under spin-transfer torque (STT). Consequently, it can then be anticipated that DMI may affect the landscape of stable states and the reversal mechanisms, which are critical to the operation of MRAM cells. In this Letter, we aim to analyze the influence of DMI on MRAM cells with perpendicular magnetization, in the range of DMI magnitude that may exist in typical material stacking used for MRAM elements. 

DMI describes the chiral exchange interaction that favors rotations between neighboring spins \cite{Dzyaloshinsky1958,Moriya1956}. The energy of an interfacial DMI between two neighboring spins $S_1$, $S_2$ can be written as 
\begin{equation}
	E_{\rm DM} = \vec{d}_{12} \cdot (\vec{S}_1 \times \vec{S}_2)\quad ,
\end{equation}
where $\vec{d}_{12}$ is the DMI vector for these spins. For an interface between perfectly isotropic films, $\vec{d}_{12}$ is given by $d\cdot \hat{e}_z \times \vec{r}_{12}$, where $d$ is the atomic DMI magnitude, $\hat{e}_z$ the unit vector normal to the interface, and $\vec{r}_{12}$ the unit vector pointing from $S_1$ to $S_2$. In the micromagnetic approximation of continuous magnetization, the interfacial DMI can be written as a volume energy density \cite{Sampaio2013}:
\begin{equation}
	\mathcal{E}_{\rm DM} = D ( m_z \cdot \partial_x m_x - m_x \cdot \partial_x m_z + m_z \cdot \partial_y m_y - m_y \cdot \partial_y m_z )\ ,
\end{equation}
where $D = C d / (a t)$ is the micromagnetic DMI magnitude, $C$, $a$ and $t$ are a geometric factor dependent on the film stacking, the lattice constant, and the thickness of the ferromagnetic film, respectively.

\begin{figure*}
\includegraphics[width=14cm]{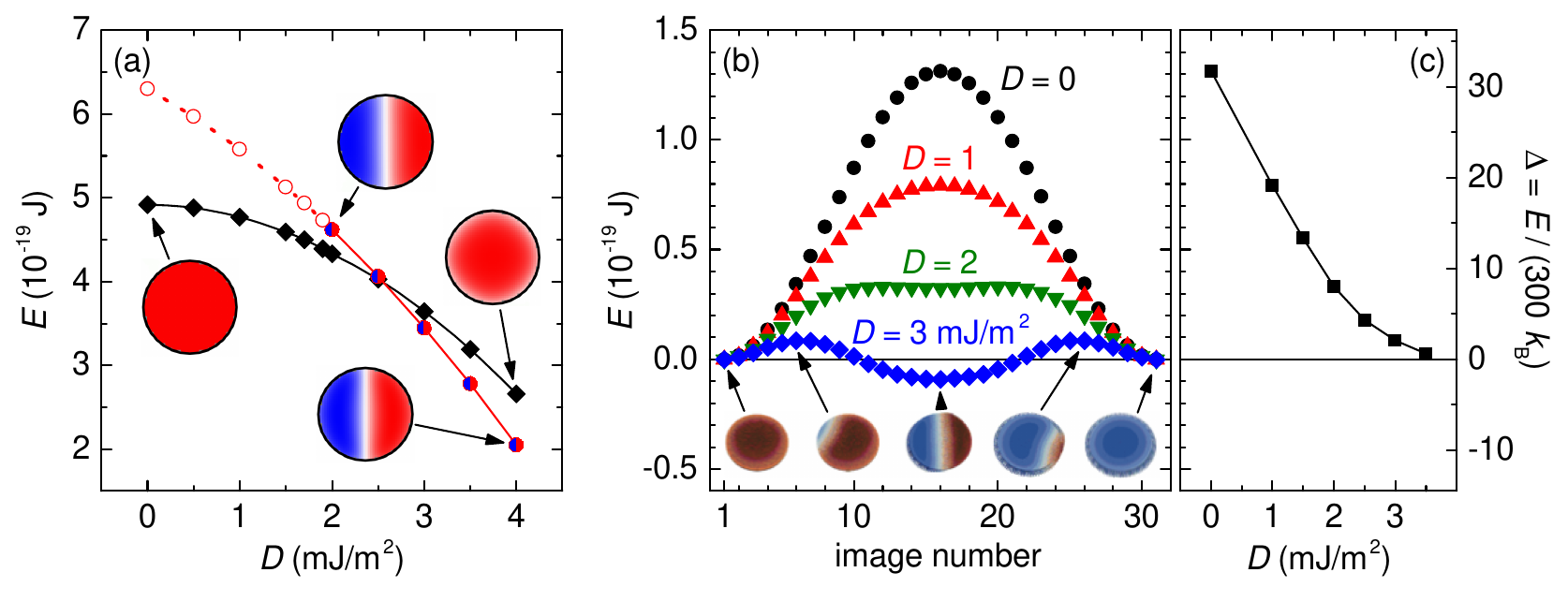}
\caption{Energy of static states in a nanodisk with DMI. (a) Energy of the quasi-uniform state (red line, circles) and of the DW state (black line, diamonds) versus DMI magnitude $D$, determined by micromagnetic simulations. The inset images show some of the stable and metastable configurations (where the colors red/white/blue correspond to the z magnetization component). The open/filled circles denote (meta)stable/unstable DW states. (b) Minimum energy paths of magnetic reversal for $D=0-3$\,mJ/m$^2$ calculated with NEB, showing the DW-mediated reversal. (c) Barrier height calculated from (b). The right axis shows the corresponding value of the room temperature thermal stability factor $\Delta$. 
}
\label{Fig1}
\end{figure*}

The DMI magnitude in thin magnetic films similar to those used in MRAM structures may reach up to a few mJ/m$^2$.\cite{Franken2014,Ryu2014} For example, recent measurements showed that $D = 0.053$\,mJ/m$^2$ for Ta/CoFe~0.6nm/MgO,\cite{Emori2014a} $1.2$\,mJ/m$^2$ for Pt/CoFe~0.6nm/MgO,\cite{Emori2014a} and $7$\,mJ/m$^2$ for Ir/Fe monolayer.\cite{Heinze2011} As we show below, even for $D$ in a range $0.3-1$\,mJ/m$^2$ we see a considerable impact on the MRAM cell performance. Performance of an STT-MRAM cell is characterized by two key parameters: the thermal stability factor $\Delta$, and the critical switching current density $j_{c0}$.\cite{Khvalkovskiy2013} $\Delta$ equals to the energy barrier height between the two magnetic states $E_b$ normalized for the operating temperature $\Delta=E_b / (k_BT)$, where $k_B$ is the Boltzmann constant; it defines the information retention time as $ t_0 \exp(\Delta)$, where $t_0$ is typically of the order of $1$\,ns. $j_{c0}$ is the zero-temperature instability threshold current density, which defines the scale of the currents required for read and write operations. In our study, we investigate how $\Delta$ and $j_{c0}$ change in presence of strong DMI effect using micromagnetic simulations. We exploit three numeric techniques: static and dynamic micromagnetic simulations using Mumax3 \cite{Vansteenkiste2014} and OOMMF \cite{Donahue1999} (for preliminary studies at $T=0$) open source codes, and nudged-elastic band (NEB) simulation of switching paths, using the FastMag code\cite{Chang2011}. We use as a model system a perpendicularly magnetized disk of $32$\,nm diameter and $1$\,nm thickness, with the following material parameters: saturation magnetization ($M_S$) of $1.03$\,MA/m, exchange stiffness ($A$) of $10$\,pJ/m, perpendicular magnetocrystalline anisotropy ($K_u$) of $0.770$\,MJ/m$^3$, and a Gilbert damping factor ($\alpha$) of $0.01$. These parameters are typical of a perpendicularly magnetized CoFeB active layer in an MTJ used in an MRAM cell. With these values, we get an effective anisotropy for the disk $K_{\rm eff} =K_u - \frac{1}{2} (N_z - N_x) \mu_0 M_s^2 = 187$\,kJ/m$^3$ (where $N_i$ are the demagnetization factors of the disk \cite{Chen1991}), corresponding to $\mu_0 H_{K_{\rm eff}} =364$\,mT, a threshold DMI $D_c = 1.7$\,mJ/m$^2$, and an $\Delta = K_{\rm eff} V / (k_BT) = 36$, calculated in a uniform rotation approximation. 

We first analyze how DMI affects the equilibrium quasi-uniform states. In these simulations made using the MuMax3 code (version 3.6.1), the magnetization was initially set up and let to completely relax. Once $D$ increases, we see that DMI induces a radial tilt of the magnetization on the borders of the disk. As a result, the total micromagnetic energy (the sum of exchange, dipolar, anisotropy, and DMI energies) reduces with $D$, Fig.\ref{Fig1}a. This observation is in agreement with other theoretical results reported for similar systems \cite{Rohart2013,Sampaio2013}. 

Next, we study the evolution with $D$ of the system energy once the magnetic disk has a straight DW in the middle, $E_{\rm DW}$, see Fig.\ref{Fig1}a. In this simulations, the magnetization distribution was generated manually. (For metastable states, the system relax in the illustrated states, the values for the unstable states were obtained using an ideal straight wall.) Even though this is not a true relaxed state, since it is symmetric it represents an energy extremum state on a possible switching path. We observe that the DMI lowers $E_{\rm DW}$ and stabilizes a N\'eel domain wall even if we started from a Bloch wall (for $D\geq0.05$\,mJ/m$^2$). The rate of variation of $E_{\rm DW}$ with $D$ follows closely the theoretical value of $-\pi S$ ($= -10^{-16} {\rm J}/({\rm J m}^{-2})$), where $S$ is the DW surface.\cite{Rohart2013} For low $D$, the DW state has a higher energy than that of a uniform state and is unstable (open circles in Fig.\ref{Fig1}a). But for $D>1.8$\,mJ/m$^2$ the DW state becomes meta-stable, which means that a DW may by trapped in the disk center if it gets there. For even larger $D$ ($D > 2.6$\,mJ/m$^2$), DW energy becomes lower than the energy of the uniform state, thus it becomes the system ground state. This is an important result as this meta-stable DW state may force the use of higher writing currents, and impairs completely the required binary operation of a typical MRAM cell (as the system no longer has only two stable states).

	
From Fig.\ref{Fig1}a we see that the energy difference between the uniform and DW state diminishes with $D$. To accurately estimate the dependence of the energy barrier on $D$, we exploit the NEB simulations \cite{Dittrich2002,Tudosa2012} implemented in the FastMag code. NEB is a method to calculate a minimum energy path (MEP), i.e. the path in a configurational space connecting two ground states (up and down states for our disk) with a trajectory having minimum energy span. Using this method, it was shown recently that for PMA MRAM cells of sizes of even 20\,nm or less, the domain-wall switching rather than the uniform rotation may be the primary thermal switching mechanism.\cite{Khvalkovskiy2013}

In Fig.\ref{Fig1}b, we show the MEP calculated using the NEB method for $D$ between $0$ and $3.5$\,mJ/m$^2$, showing the intermediate magnetic states as insets. These simulations show that MEP is the DW-mediated reversal for all considered values of $D$ ($0 - 3.5$\,mJ/m$^2$), confirming the qualitative conclusion from Fig.\ref{Fig1}a. It also confirms existence of the metastable states for the DW for $D\gsim2$\,mJ/m$^2$ (corresponding to the appearance of an intermediate energy minimum in the curve of Fig.\ref{Fig1}b). For larger $D$ ($D>2.5$\,mJ/m$^2$), DW at the center of the disk becomes the ground state, and the highest energy point on MEP becomes an intermediate DW state close to an edge (see the magnetization distribution in the insets to Fig.\ref{Fig1}b).

The calculated by NEB simulations energy barrier $E_B$ and $\Delta$ as a function of $D$ obtained with these NEB simulations is plotted in Fig.\ref{Fig1}c. For $D$ = 0 we get $\Delta = 33$, which is close to the analytical result $\Delta = 36$. This shows that without DMI, the energy difference between the uniform rotation and DW-mediated reversal is small, and DMI strongly promotes the DW-mediated reversal. Once $D$ increases, $\Delta$ drops dramatically even for moderate values of $D$: for $D=0.5$\,mJ/m$^2$, $\Delta$ drops by 20\% to 27, and for $D=1$\,mJ/m$^2$, $\Delta$ drops by 40\%, to 20, with the corresponding reduction of the retention time by six orders of magnitude. For even larger $D$, we see that the barrier vanishes completely.   

We now investigate the effects of DMI on the STT-induced switching performance. First, we simulate STT switching of our FL at zero temperature. We assume spin polarization $P$ of $40$\%, with in-plane torque $\vec{m}\times \vec{p}\times \vec{m}$, $\vec{p}$ being the orientation of the injected spins, and no out-of-plane torque\cite{Slonczewski1996,Berger1996}. The simulations show that the switching process starts by an excitation of oscillations that increase in amplitude, until magnetization breaks into the two-domain state with  a subsequent reversal of the disk by a DW propagation. For $D=0$, the amplitude of the oscillations increases gradually and uniformly in the disk, while, for finite $D$, the oscillations are uneven in amplitude, and strongly localized at the border of the disk. This may make the reversal process quite sensitive on the border properties, such as its shape and roughness, but also on the spatial discretization of the simulation (see supplementary materials \cite{SupplMat}). To avoid the artefacts related to the boundary discretization, we used the FastMag code in these simulations; its finite element micromagnetic solver allows  defining the simulated disk with a smooth border.

\begin{figure}
\includegraphics[width=7cm]{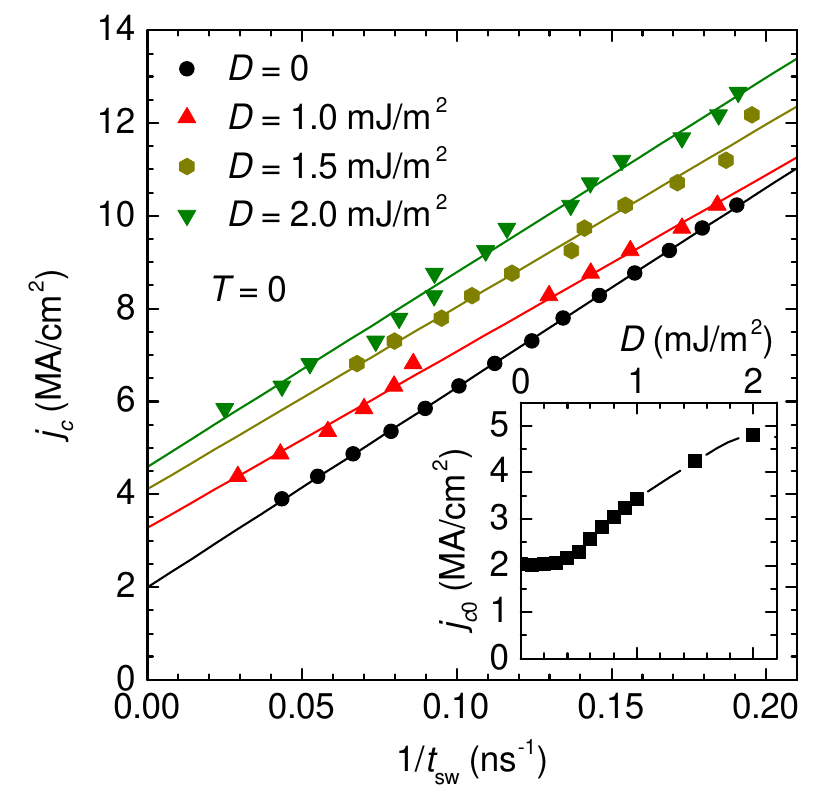}
\caption{Switching under current (STT) at zero temperature. Simulated applied current versus reciprocal switching time for different $D$ values. The lines are linear fits to Eq.\ref{eqtswitch}. The inset plot shows the extracted $j_{c0}$ as a function of $D$.}
\label{Fig2}
\end{figure}

For each value of the current density $j$ we extract the switching time $t_{\rm sw}$ of our FL, defined as the time when the FL magnetization crosses the equatorial plane (plane $z$ = 0). In Fig.\ref{Fig2}a, we show the simulation result for $1/t_{\rm sw}$ as a function of $j$, for $D$ ranging between $0$ and $2$\,mJ/m$^2$. We find that the switching time at a given current density is always larger for larger $D$. For an MRAM cell, the FL switching time $t_{\rm sw}$ varies inversely with $j$ as follows \cite{Sun2000}:
\begin{equation}
t_{\rm sw}^{-1} \propto  j/j_{c0}-1\quad.
\label{eqtswitch}
\end{equation}
We use Eq.\ref{eqtswitch} to fit the switching data and extract $j_{c0}$. It appears that even for large $D$ the data is reasonably linear in $j$, which allows us to fit this data using Eq.\ref{eqtswitch}. The fit result, $j_{c0}$, is shown in as a function of $D$ in the inset to Fig.\ref{Fig2}. For $D=0$ we get $j_{c0}$=$2$\,MA/cm$^2$. For $D=0.5$\,mJ/m$^2$, $j_{c0}$  increased by already 15\%, reaching $2.3$\,MA/cm$^2$. For larger $D$, it increases at a striking pace, reaching $3.4$\,MA/cm$^2$ (70\% increase) for $D=1$\,mJ/m$^2$ and $4.2$\,MA/cm$^2$ (110\% increase) for $D=1.5$\,mJ/m$^2$. For even higher values of $D$ ($>2$\,mJ/m$^2$), the system reaches often metastable states (with a DW), which impedes the determination of switching times.

As we mentioned above, DMI promotes switching via very non-uniform modes. Consequently, the cell switching performance and it dependence on $D$ may become sensitive to the shape of the sample. In order to verify this suggestion, we perform additional simulations of the STT switching of the cells with different shapes. We find that while for $D=0$, $j_{c0}$ does not depend much on the cell, for finite $D$ this dependence is considerable. For instance,  $j_{c0}$ for $1$\,mJ/m$^2$ ranged from $3.3$ up to $8.3$\,MA/cm$^2$. These findings support the importance of border resonant modes in the reversal process in the presence of DMI \cite{Kim2014}. See supplementary materials for more information about the study of the role of edges and the dynamics of the switching \cite{SupplMat}. 
 
\begin{figure}
\includegraphics[width=7cm]{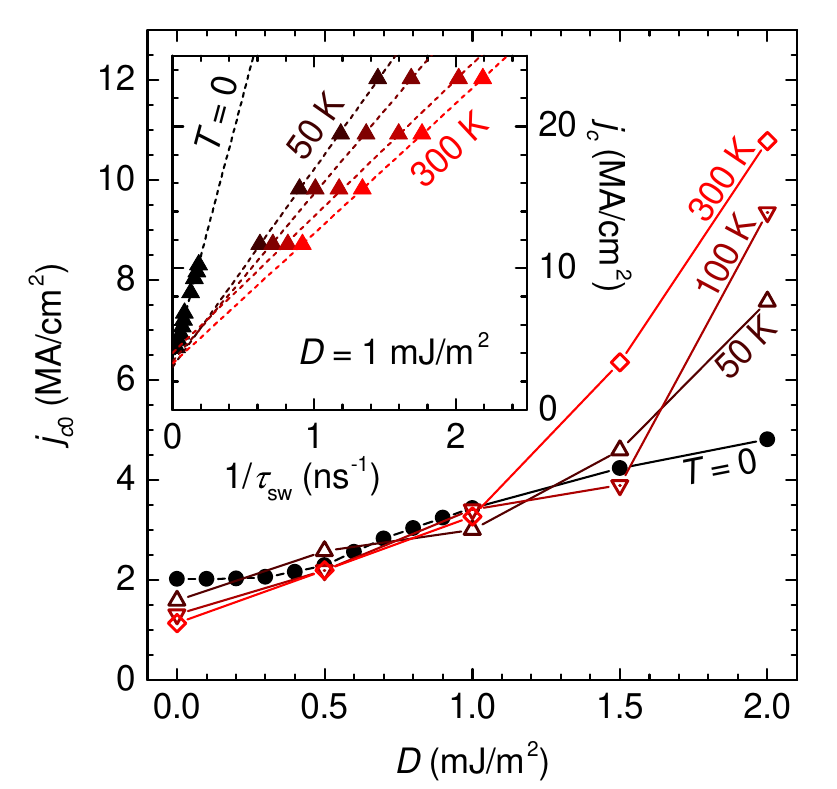}
\caption{Effects of DMI on the thermal stability and current induced switching of MRAMs. $j_{c0}$ versus $D$ for $T=0$, $50$, $100$, and $300$\,K. 
The inset plot is the current versus the reciprocal mean switching time ($\tau_{\rm sw}$) for $D=1$\,mJ/m$^2$, for temperatures of 50, 100, 200 and 300\,K, extracted from multiple ($60$ to $80$) stochastic simulations; the data for $T=0$ are also shown.}
\label{Fig3}
\end{figure}

We see that DMI leads to an increase of critical switching current ($j_{c0}$) with simultaneous decrease of the thermal stability factor ($\Delta $). These opposing effects suggests that switching with STT at finite temperature might be very different from the $T=0$\,K case that we calculate in Fig.\ref{Fig2}. To take the thermal effects and DMI into account in determining $j_{c0}$, we performed stochastic dynamical simulations, where we introduced a random magnetic field with a Gaussian amplitude distribution to simulate the effects of temperature \cite{Vansteenkiste2014}. We simulated repeatedly (at least twenty times) a current pulse with the same STT parameters as before for each set of parameters ($D$, $j$, and $T$), and calculated the mean switching time $\tau_{\rm sw}$. In the inset of Fig.\ref{Fig3}, we show $j$ versus $1/\tau_{\rm sw}$ at $D=1$\,mJ/m$^2$ for various values of temperature. We extrapolated $j_{c0}$ as before.

In Fig.\ref{Fig3}, we show the variation of $j_{c0}$ with $D$ for various values of the temperature. We observe that $j_{c0}$ always increases with $D$, with this increase being larger for higher $T$. The rise of $j_{c0}$ is exacerbated by temperature: while at $0$\,K the $j_{c0}$ at $D=2$\,mJ/m$^2$ is twice that of $D=0$, at $300$\,K the difference is fivefold. For $D=0$, we see that $j_{c0}$ decreases for higher temperature. This result is in agreement with the stochastic macrospin simulations, which also show that even in a uniform switching mode and with a great statistical quality $j_{c0}$  is expected to decrease with the temperature (see supplementary materials for details \cite{SupplMat}). However for large $D$, we see that this dependence is reversed, and $j_{c0}$ becomes larger for larger $T$. 

Finally, the influence of DMI on both the MRAM switching current and thermal stability, quantified by $j_{c0}$ and $\Delta$, can also be seen in Fig.\ref{Fig3} and Fig.\ref{Fig1}c.
We see readily that even a moderate DMI of $D\sim0.5$\,mJ/m$^2$ leads to an increase of $j_{c0}$ and a large decrease of the thermal stability by tens of percent.  This result emphasizes the importance of quantification and minimization of the DMI magnitude in materials used for the free layers in MRAM cells, possibly using materials that induce DMI of opposing sign\cite{Hrabec2014}.

During the preparation of this article, an article by Jang {\it et al.} appeared \cite{Jang2015}, which discusses some of the points also included here.

\begin{acknowledgments}
This work was supported by the Samsung Global MRAM Innovation Program, and by the NSF grants DMR-1312750 and CCF-1117911.
\end{acknowledgments}

%

\end{document}